\documentstyle[12pt,psfig]{article}

\newcommand{\rf}[1]{(\ref{#1})}
\newcommand{\bea}{\begin{eqnarray}}
\newcommand{\eea}{\end{eqnarray}}

\newcommand{\e}{\mbox{e}}

\newcommand{\g}{\gamma}

\renewcommand{\l}{\lambda}

\renewcommand{\b}{\beta}

\newcommand{\n}{\nu}

\newcommand{\Del}{\Delta}
\newcommand{\sg}{\sigma}

\newcommand{\oh}{\frac{1}{2}}

\newcommand{\ra}{\right\rangle}
\newcommand{\la}{\left\langle}

\newcommand{\mi}{\!-\!}
\newcommand{\equ}{\!=\!}
\newcommand{\pl}{\!+\!}

\newcommand{\no}{\nonumber}

\def\void{}
\def\labelmark{}

\newenvironment{formula}[1]{\def\labelname{#1}
\ifx\void\labelname\def\junk{\begin{displaymath}}
\else\def\junk{\begin{equation}\label{\labelname}}\fi\junk}%
{\ifx\void\labelname\def\junk{\end{displaymath}}
\else\def\junk{\end{equation}}\fi\junk\labelmark\def\labelname{}}

{\ifx\void\labelname\def\junk{\end{array}\end{displaymath}}
\else\def\junk{\end{array}\right.\end{equation}}
\fi\junk\labelmark\def\labelname{}\def\junk{}
}

\newcommand{\beq}{\begin{formula}}
\newcommand{\eeq}{\end{formula}}
\newcommand{\beqv}{\begin{formula}{}}

\begin{document}

\hfill NBI-HE-99-15

\hfill AEI-100


\hfill 2 Apr 1999

\begin{center}
\vspace{24pt}
{ \Large \bf A New Perspective on Matter Coupling \\
\hspace{3cm} \\
in 2d Quantum Gravity}

\vspace{48pt}

{\sl J. Ambj\o rn,}$\,^{a,}$\footnote{email ambjorn@nbi.dk}
{\sl K. N. Anagnostopoulos}$\,^{a,}$\footnote{email konstant@nbi.dk}
and
{\sl R. Loll}$\,^{b,}$\footnote{email loll@aei-potsdam.mpg.de}

\vspace{24pt}

$^a$~The Niels Bohr Institute, \\
Blegdamsvej 17, DK-2100 Copenhagen \O , Denmark 

\vspace{24pt}
$^b$~Max-Planck-Institut f\"{u}r Gravitationsphysik,\\
Albert-Einstein-Institut,\\
Am M\"uhlenberg 5, D-14476 Golm, Germany

\vspace{36pt}

\end{center}

\vspace{1.5cm}

\begin{center}
{\bf Abstract}
\end{center}

\vspace{12pt}
\noindent
We provide compelling evidence that a previously introduced model of
non-pertur\-ba\-tive 2d Lorentzian quantum gravity exhibits 
(two-dimensional) flat-space behaviour when coupled to Ising spins. 
The evidence comes from both a high-temperature expansion 
and from Monte Carlo simulations of the combined gravity-matter 
system. This 
weak-coupling 
behaviour lends further support to the 
conclusion that the Lorentzian model is a genuine alternative to 
Liouville quantum gravity in two dimensions, with a different, and
much `smoother' critical behaviour. 
 

\newpage

\section{Introduction}\label{intro}

At the end of the twentieth century, the non-perturbative quantization
of gravity remains an elusive goal for theoretical researchers.
There is not even a consensus on how the problem should best be tackled.
For example, considering pure-gravity approaches, we have on the one hand 
Euclidean path-integral methods, which are close to usual formulations 
of (non-generally covariant) 
quantum field theories and well-suited for numerical simulations. 
On the other hand, in canonical quantization approaches it is -- at
least in principle --
easier to address questions about the behaviour of spatial 
three-geometries, but the complicated structure of the constraints 
tends to lead to computational difficulties.\footnote{{Alternatively, 
one could embed quantum gravity in a larger, unified theory like 
string theory or (the as yet non-existent) M-theory. However, these
are still far from giving us any detailed information about 
the quantum gravity sector.}} 
Unfortunately, very little is known about the relation between the
covariant and canonical approaches. In part this is due to 
the `signature problem' of the path-integral formulations:
the sum over all space-time geometries is usually taken over
Riemannian, and not over the physical Lorentzian (pseudo-Riemannian) 
four-metrics modulo diffeomorphisms. The problem of how to relate
the two sectors by an appropriately generalized Wick rotation
remains unresolved.

Our aim is to investigate the possible consequences of taking the 
Lorentzian structure seriously within a path-integral approach.
In order to gauge the difficulties this involves and 
to circumvent technical problems, we first addressed the
issue in two space-time dimensions, where there already exists a 
well-understood theory of (Euclidean) quantum gravity, namely,
Liouville gravity. In \cite{al}, we proposed a new, Lorentzian
model of 2d quantum gravity, obtained by taking the continuum
limit of a state sum of
dynamically triangulated two-geometries. The Lorentzian aspects
of the model were two-fold: firstly, the sum was taken only over those 
two-geometries which are generated by evolving a one-dimensional
spatial slice and which 
allow for the introduction of a causal structure. Secondly, 
the Lorentzian propagator was obtained by a suitable analytic
continuation in the coupling constant. The first aspect turned out to
be the crucial one, leading to a continuum theory of 2d quantum 
gravity {\it inequivalent} to the usual Liouville gravity.
This was shown in \cite{al}, where both the loop-to-loop
propagator and various geometric properties of the model were
calculated explicitly. The Hausdorff dimension of the quantum
geometry is $d_{H}\equ 2$, and points to a much smoother behaviour
than that of the Euclidean case (where $d_{H}\equ 4$). 

\begin{figure}
\centerline{\hbox{\psfig{figure=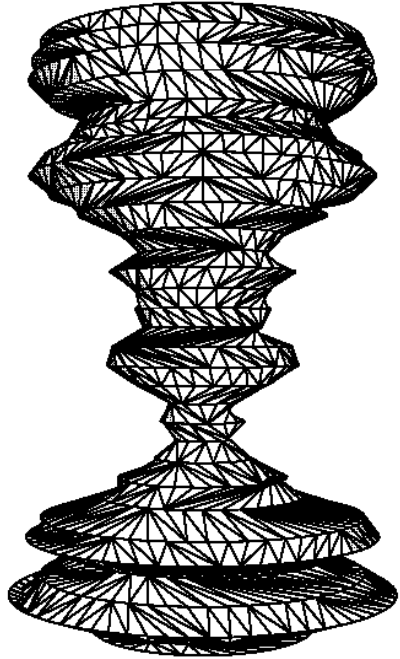,height=10cm}}}
\caption[fluctuate]{A typical discrete history of pure Lorentzian gravity
with volume N=1024.}
\label{fluctuate}
\end{figure}

However, we must emphasize that $d_H \equ 2$ does {\it not} imply 
a {\it flat} geometry.
The model of Lorentzian gravity defined in \cite{al} allows 
for arbitrarily large fluctuations of the spatial volume from
one time-slice to the next.
This is illustrated by Fig.\ \ref{fluctuate}, which
shows a typical surface generated by the Monte Carlo simulations,
to be described in Sec.\ \ref{simulation}. 
The length of the compact spatial slice 
fluctuates strongly with time (pointing along the vertical axis).
Using the results of \cite{al}, one easily derives 
that in the thermodynamic limit and for large times
the average spatial volume $L$ and fluctuations around $L$ behave like 
\beq{add1.1}
\la L \ra = \frac{1}{\sqrt{\Lambda}}~~~~{\rm and}~~~~
\la \Delta L \ra = 
\sqrt{\la L^2\ra - \la L\ra^2} = \frac{1}{\sqrt{2\Lambda}}
\eeq
respectively, for a given cosmological constant $\Lambda$. 
This demonstrates that even in the continuum limit 
the fluctuations are large, and of the same order of magnitude 
as the spatial volume itself.
 
We managed in \cite{al} 
to further trace the difference between the two quantum theories
to the presence or absence of so-called baby universes. 
These are outgrowths of the two-geometry giving it the structure
of branchings-over-branchings, which are known to dominate the
typical geometry contributing to the Euclidean state sum. 
On the other hand, in the Lorentzian state sum, one can suppress
the formation of such branchings {\it with respect to the
preferred spatial slicing} (which is not present in the Euclidean
picture, where no directions are distinguished). There is also
a physical motivation for suppressing the generation of baby
universes, since the associated (discrete) geometries can
usually not be embedded isometrically in a smooth Lorentzian 
space-time. If nevertheless one {\it did} decide to generalize the
evolution rules of the Lorentzian model to allow for such
branchings (and keep only a weaker notion of causality, c.f. \cite{al}),
one would rederive the usual Euclidean Liouville results. 
In what follows, when talking about `the Lorentzian model', we will
mean the unmodified model without branching baby universes, i.e. the
model of 2d quantum gravity that does not lie in the same universality
class as Liouville gravity. We also would like to point out that
from the point of view of a canonical 
quantization the Lorentzian model is much more natural. The 
inclusion of topology changes of space into a canonical scheme 
would require a so-called third quantization of geometry.  

In Liouville gravity, matter couples strongly to geometry, perhaps
even too
strongly in the sense that the combined system becomes inconsistent
when the central charge of the (conformal) matter exceeds one. 
Arguments have been presented which link the strong deformation of 
geometry to the creation of baby universes \cite{adj}. It is therefore 
conceivable that the Lorentzian model of gravity -- where baby
universes are absent -- has a weaker and less
pathological coupling to matter.

In order to understand the behaviour of the combined
gravity-matter system, we are considering here the coupling
of the gravitational model of reference \cite{al} 
to an Ising model of spin-$\frac{1}{2}$, with nearest-neighbour
interaction $\sum_{<ij>}\sigma_{i} 
\sigma_{j}$ between its spins $\sigma_{i}\equ\pm 1$.
In \cite{al} we made a careful analysis of the implications
of the Lorentzian signature for the sum over space-time metrics. 
The most straightforward way of obtaining
the continuum limit consisted in performing the calculations
in the discretized model with purely imaginary coupling
(corresponding to Euclidean signature) and  
only afterwards `rotating back' to the Lorentzian sector.
Moreover, it turned out that certain simple properties, like the
fractal dimension of space-time, were independent of the
analytic continuation. 

We will apply the same philosophy in the present context
by analyzing the Ising model coupled to 2d gravity with
coupling constants corresponding to the Euclidean signature
sector. 
Nevertheless we will continue to talk about `Lorentzian' gravity 
coupled to matter, because the choice 
of two-dimensional Euclidean geometries contributing to 
the path integral is dictated by the requirement that 
after the rotation to Lorentzian signature
they should be causal and non-singular. 
The `matter observables' we will consider are the 
critical exponents for the Ising model, 
characterizing the underlying $c\equ 1/2$ fermionic continuum  
model coupled to gravity, which are not expected to change 
under the rotation to Lorentzian signature. 
In order to determine 
the universality class of the interaction between matter and 
gravity it is therefore convenient to work 
entirely within the Euclidean sector of our Lorentzian
gravity model. 

For fixed regular two-dimensional lattices, and in the absence of
an external magnetic field, the Ising model can be solved exactly in
a variety of ways (see, for example, \cite{nm,mw,dombo1}). The
partition function (for the square lattice) was found by
Onsager. Its critical behaviour is characterized 
by a logarithmic singularity of the specific heat and the
critical exponents near the Curie temperature, $\alpha\equ 0$,
$\beta\equ 0.125$, and $\gamma\equ 1.75$, for the specific heat, the
spontaneous magnetization and the susceptibility respectively. 

For the case of the usual Euclidean 2d gravity, described by an
ensemble of planar random surfaces, coupling to Ising spins was
first considered in \cite{kazakov}, where an exact solution was
obtained with the help of matrix model methods. 
It could be shown that in the presence of gravity, the matter 
behaviour is `softened' to a third-order phase transition,
characterized by critical exponents $\alpha\equ -1$, $\beta\equ 0.5$, and
$\gamma\equ 2$ \cite{bk}. On the other hand, the geometry is 
`roughened', as exemplified by the increase 
from $-1/2$ (pure 2d Liouville gravity)
to $-1/3$ of the entropy exponent $\g_{string}$ 
for baby universes on manifolds of spherical topology. 

It is not entirely straightforward to apply the 
methods used to obtain these exact solutions to the Lorentzian 
gravity model. For example, one can write down an expression for the
transfer matrix generalizing that of the Onsager solution by imposing
a length cutoff $l_{0}$ on the length of spatial slices. 
However, a major stumbling block to understanding the behaviour of
its eigenvalues as $l_{0}\rightarrow\infty$ is the fact that as a
consequence of the gravitational degrees of freedom, transitions
between spatial slices of different length are allowed. 
This makes the use of Fourier transforms problematic, which are
an essential ingredient of this and other algebraic solution schemes. 
Moreover,
the Hilbert space dimension for the discrete, finite model is
given by $\sum_{l=1}^{l_{0}}2^{l}$, which grows rapidly with $l_{0}$. 

In the absence of an analytic exact solution\footnote{A matrix model of 
Lorentzian gravity coupled to Ising spin has
been formulated recently. Its analysis is the subject of a 
forthcoming publication \cite{xxx}.}, one way to try to extract
information about the matter-coupled model is by performing  
a series expansion of the partition function $Z$ at high or low temperature,
or of suitable derivatives of $Z$. For flat, regular lattice
geometries, these have been studied extensively since the early days of
the Ising model. It is well-known that the high-$T$ expansion, in
particular, that of the magnetic susceptibility $\chi$ at zero field
is well-suited for obtaining information about the critical
behaviour of the theory. We will show that the same is true for
the coupled gravity-Ising model, after taking into account some
peculiarities to do with the fact that we have an ensemble of fluctuating
geometries instead of a fixed lattice. In the limit of large lattice size 
$N$, there is a well-defined expansion in terms of $u:=\tanh \beta $,
where the coupling $\beta$ is proportional to the inverse temperature,
whose coefficients can be determined by diagrammatic techniques. 
Given a plausible ansatz for the singularity structure of the
thermodynamic functions, one can then extract estimates for the
critical point and critical exponents from the first few terms of such
an expansion. These results are corroborated 
by performing a Monte Carlo simulation of 
Lorentzian gravity coupled to the Ising model. Apart from 
being in good agreement with the high-$T$ expansion, the 
simulations also allow us to measure the quantum geometrical properties 
of the model.

\section{The high-$T$ expansion}\label{expansion}

Recall the usual high-$T$ expansion of the Ising model on a fixed
lattice of volume $N$, with partition function
\beq{2.1}
Z(\beta,N) =\sum_{ \{\sigma_i =\pm 1\} } \e^{\b
\sum\limits_{<ij>}
\sigma_i\sigma_j + H\sum\limits_{i} \sigma_{i}},
\hspace{2cm} \beta=\frac{J}{kT},
\eeq
where the sum is taken over all possible spin configurations, and
$J>0$ denotes the ferromagnetic Ising coupling. We will only consider
the case of vanishing magnetic field, $H\equ 0$. The Ising spins are located at
the lattice vertices, labelled by $i,j\in 1\ldots v$. 
A convenient expansion parameter at high temperature is $u:=\tanh \beta$,
which we can use to re-express
\beq{2.2}
\e^{\b\sigma_i \sigma_j}=(1+u\, \sigma_i \sigma_j) 
\cosh \b.
\eeq
Substituting \rf{2.2} into \rf{2.1}, the partition function becomes
\bea
Z(\beta,N)&=&(\cosh \b )^s \sum_{\{\sigma_i \} }
\Bigl[ 1+u\sum_{<ij>}
\sigma_i\sigma_j +u^2 \sum_{<ij>} \sum_{<kl>} (\sigma_i\sigma_j)
(\sigma_k\sigma_l)+\ldots\Bigr]\label{2.3} \\
&=:& 2^v (\cosh \b )^s (1+\sum_{n\geq 1} \Omega_n u^n ),
\label{2.4}
\eea
with $v$ denoting the number of vertices and $s$ the number of
nearest-neighbour pairs (i.e. the number of lattice links). Note that
the terms $\sim u^n$ in eq.\ \rf{2.3} are only non-vanishing if every 
$\sigma_i$ in $\sigma_{i_1} \sigma_{i_2}\ldots \sigma_{i_n}$ appears
an even number of times. Representing spin pairs $(\sigma_i \sigma_j)$
by drawing a link between $\sigma_i$ and $\sigma_j$ on the lattice,
this is equivalent to the following statement: non-vanishing contributions
to $\Omega_n$ in eq.\ \rf{2.4} 
correspond to figures of lattice links which are closed
polygons, with an even number of links meeting at each vertex.
The coefficient $\Omega_n$ simply counts the number of such figures at order $n$ 
that can be put down on a given lattice, and will depend on the lattice
geometry (triangular, square, etc.). It is a polynomial in the 
variable $N$.

Because of the extensive nature of the free energy $F(N)\equ -kT \ln Z(N)$,
we must have that $(1+\sum \Omega_n u^n )\sim \e^{N (\ldots )}$ in the
thermodynamic limit $N\to\infty$, and we can therefore 
write for the partition function per unit volume
\beq{2.5}
\ln Z(\beta):=\frac{1}{N} \ln 
Z(\beta,N)=\frac{s}{N}\, \cosh\b +\frac{v}{N}\, \ln 2 
+\sum_{n\geq 1}\omega_{n}^{(0)}u^{n},
\eeq
where $\omega_{n}^{(0)}$ is obtained by taking the term linear in $N$
in $\Omega_{n}$ and setting $N\equ 1$. Note that both connected and
disconnected graphs contribute to $\omega_{n}^{(0)}$.
A similar relation can be obtained
for the magnetic susceptibility at zero field, $\chi (N)=k 
T\frac{\partial^{2}}{\partial H^{2}}\ln Z(N)|_{H=0}$. 
At high temperature, the
susceptibility per unit volume can be expressed as
\beq{2.6}
\chi=k T\ (1+\sum_{n\geq 1}\omega_{n}^{(2)}u^{n}).
\eeq
The coefficients $\omega_{n}^{(2)}$ are the exact analogues of
$\omega_{n}^{(0)}$ in eq.\ \rf{2.5}, where the counting now refers to
polygon graphs with two odd vertices (vertices with an odd number of 
incoming links), and all other vertices even 
(c.f. \cite{dombo2}, but beware of the
difference in notation for the number of vertices).

Since we are primarily interested in the bulk behaviour of the
gravity-matter system, we will in the following for simplicity 
choose the boundary conditions to be periodic. That is, we will identify
the top and bottom spatial slices of the cylindrical histories introduced
in \cite{al}. Clearly this is not going to affect the local properties
of the model. As above, we will denote the discrete volume,
i.e. the number of triangles of a given two-dimensional geometry (with
torus topology), by $N$. It follows immediately that such a geometry
contains $N$
time-like links, $N/2$ space-like links, $N/2$ vertices and $3 N/2$
nearest-neighbour pairs. 

In quantum gravity the volume $N$ becomes a dynamical variable.
For fixed topology, the only coupling constant appearing in the action 
of pure 2d quantum gravity 
is the cosmological constant, multiplying the volume term.
The partition function of the Ising model coupled 
to 2d Lorentzian quantum gravity is given by
\beq{add2.1}
G(\l,t,\b) = \sum_{T \in {\cal T}_t} e^{-\l N_T} Z_T(\b)=
\sum_{T \in {\cal T}_t} e^{-\l N_T} 
\sum_{ \{\sigma_i(T)\} } \e^{\b
\sum\limits_{\la ij\ra \in T}
\sigma_i\sigma_j },
\eeq
where the sum is taken over all triangulations $T$ with the topology
of a torus and $t$ time-slices, $N_T$ is the number of triangles 
in $T$, and $Z_T(\b)$ the Ising partition function
\rf{2.1} defined on $T$. Fortunately, the summation
over volumes in eq.\ \rf{add2.1} does not lead to additional 
complications in the analysis of the thermodynamic properties
of the spin system, since the state sums for fixed and fluctuating
volume are simply related by a Laplace transformation.
Rewrite relation \rf{add2.1} as
\beq{add2.2}
G(\l,t,\b)= \sum_N e^{-\l N} \tilde{Z}(\b,N,t) :=
\sum_N e^{-\l N} \sum_{T\in {\cal T}_{N,t}} Z_T(\b),
\eeq
where ${\cal T}_{N,t}$ denotes the toroidal triangulations of
volume $N$ and length $t$ in the time direction. 
Analogous to eq.\ \rf{2.5}, we expect 
the matter part $f(\b)$ of the free energy density in the 
gravitational ensemble to behave in the thermodynamic 
limit ($N \to \infty$ and $t \propto \sqrt{N}$) 
like\footnote{Note that with the conventions used 
in definition \rf{2.1}, the ground 
state energy is $-3\b N/2$ and the free energy density $f(\b)$ is 
negative.}
\beq{add2.4}
\tilde{Z}(\b,N,t) \to e^{(\l_c-\b f(\b)) N+ o(N)}.
\eeq
(For simplicity, we have set the ferromagnetic coupling to $J\equ 1$.)
We can now reexpress eq.\ \rf{add2.2} as
\beq{add2.5}
G(\l,t,\b) = \sum_N e^{(\l_c(\b)-\l)N + o(N)},~~~~
\l_c(\b) = \l_c -\b f(\b),
\eeq
where $\l_c\equiv \l_{c}(\b \equ 0) \equ\ln 2$ is the critical cosmological 
constant of pure gravity, which was determined in \cite{al}. 
Interesting limiting cases are $\b \to 0$ where 
$ -\b f(\b) = \oh\ln 2$, reflecting 
the factor $2^v$ in eq.\ \rf{2.3} (each spin has two states),
and the strong coupling region 
$\b \to \infty$ where $-\b f(\b) \to 3\b/2$ (only the
ground state of all spins aligned contributes to 
the state sum \rf{2.1}).
The term proportional to the pure gravity cosmological 
constant $\l_c$ appearing together with the free energy in \rf{add2.4}
has its origin in the sum over all triangulations,
\beq{add2.3a}
\sum_{T\in {\cal T}_{N,t}}1 = e^{\l_c N+ o(N)}.
\eeq
A calculation of $\tilde{Z}(\b,N,t)$ not only 
determines the thermodynamic 
properties of the spin system in the presence of gravity, but at
the same time describes 
gravitational aspects of the coupled system, for example, the 
{\it critical cosmological constant} $\l_c(\b)$. Conversely, 
knowledge of $\l_c(\b)$ determines the spin partition function
in the infinite volume limit. -- 
The analogue of the high-$T$ expansion \rf{2.4} in the presence of gravity 
is given by
\beq{2.7}
\tilde{Z}(\beta,N,t)=(\cosh 
\b)^{\frac{3N}{2}}\ 2^{\frac{N}{2}}\sum_{T\in {\cal T}_{N,t}}
(1+\sum_{n\geq 1} \tilde{\Omega}_n (T) u^n ).
\eeq
We may reexpress the critical cosmological
constant of the combined system as
\beq{add2.6}
\l_c(\b)=\l_c+\frac{3}{2} \ln \cosh \b +\oh \ln 2 + \tilde{f}(u),
\eeq
where $\tilde{f}(u)$ is defined in the thermodynamic limit by
\beq{add2.8}
\frac{\sum\limits_{T\in {\cal T}_{N,t}}
(1+\sum\limits_{n\geq 1} \tilde{\Omega}_n (T) u^n )}
{\sum\limits_{T\in {\cal T}_{N,t}}1} 
= e^{N \tilde{f} (u)}.
\eeq
The coefficients $\tilde{\Omega}$ of the power series 
now depend on the triangulation $T$. 
When counting diagrams of a given type and order $n$, we must keep in
mind that the vertex neighbourhoods do not look all the same, as they
do in the case of a regular lattice, but that the distribution of
coordination numbers (numbers of links meeting at a vertex) is subject
to a probability distribution. The coefficients in the high-$T$ expansion 
therefore count the {\it average} occurrence of a certain diagram type in
the ensemble of triangulations of a fixed volume $N$, for large $N$.

\begin{figure}
\centerline{\hbox{\psfig{figure=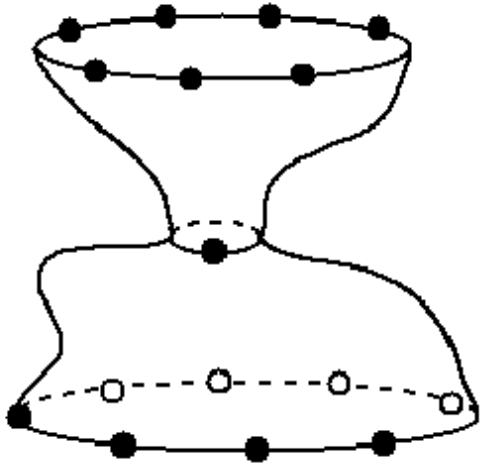,height=7cm}}}
\caption[fig1]{A two-dimensional geometry with a `pinch' of
length 1.}
\label{fig1}
\end{figure}

Starting to evaluate the series in \rf{2.7} order by order, one
immediately notices a qualitative difference from the regular case.
If we had considered a regular triangular lattice (coordination
number 6), the first non-trivial contribution to the counting of even
diagrams would have appeared at $n\equ 3$,
where one obtains $\Omega_3(N)\equ N$, coming from closed triangle graphs.
However, when looking at all two-dimensional random lattices contributing
to the sum over geometries in the gravity case, there are geometries
which have one or several `pinches'. A pinch is a spatial slice of
minimal length $l\equ 1$, which consists of a single link and a single 
vertex (see Fig.\ \ref{fig1}).
Pinches occur even if the total volume of the two-geometry is kept fixed, 
since in the presence of gravity the length of spatial slices is a 
fluctuating dynamical variable.

For the gravitationally coupled Ising model, the lowest-order contribution 
to the power series in $u$ in \rf{2.7} occurs therefore already at
order $n\equ 1$. Clearly, such pinching contributions will be present at
all orders, in both connected and disconnected diagrams, on top of the
ordinary `bulk' contributions, coming from diagrams which do not wind
around the spatial direction of the torus in a non-trivial way. The
former have no analogue on regular lattices.

Fortunately, it turns out that the pinch contributions are
irrelevant, in the sense that they contribute at a lower order
of $N$, whereas the bulk contributions in $\tilde{\Omega}_n$ go like
$N^k$, $k\geq 1$. 
This can be seen most easily by considering the Laplace-transformed
partition function. Let us begin by evaluating
the zeroth-order term of eq.\ \rf{2.7},
\beq{2.8}
G(\tilde{\lambda},t)=
\sum_N \e^{-\tilde{\lambda} N} \sum_{T\in {\cal T}_{N,t}} 1:=
\sum_N \e^{-\lambda N} (\cosh \b)^{\frac{3N}{2}}\ 
2^{\frac{N}{2}} \sum_{T\in {\cal T}_{N,t}} 1,
\eeq
where for notational brevity we have defined 
an `effective cosmological constant'
$\tilde{\lambda}\equ\lambda -\frac{3}{2}\ln \cosh \b -
\frac{1}{2} \ln 2$, in accordance with eq.\ \rf{add2.6}. 
The left-hand side of \rf{2.8} can be computed
as
\beq{2.9}
G(\tilde{\lambda},t)=\oint \frac{dx}{2\pi i x}\; G(x,y=\frac{1}{x};
\e^{-\tilde{\lambda}};t),
\eeq
given the explicit form of the propagator derived in ref.\ \cite{al}, to
which we also refer for details of notation.
The term proportional to $u^{1}$ in the Laplace transform of
\rf{2.7} is
\beq{2.10}
\sim u^{1}:\hspace{0.8cm} \sum_{N} \e^{-\tilde{\lambda} N} 
\sum_{T\in {\cal T}_{N,t}^{l=1}} 1=G(\tilde{\lambda},t)\, 
\tilde{\Omega}_1^{\rm norm}(\tilde{\lambda}),
\eeq
where the second summation is over triangulations with a single `pinch'
of spatial length $l\equ 1$. To arrive at the last expression on the 
right-hand side, the factor $G(\tilde{\lambda},t)$ has been pulled
out. In terms of quantities derived in \cite{al},
the normalized coefficient $\tilde{\Omega}_1^{\rm norm}$
is most easily computed as
\beq{2.11}
\tilde{\Omega}_1^{\rm norm}(\tilde{\lambda})=
\frac{\sum\limits_{{\tilde 
t}=1}^{t-1}G_{\tilde\lambda}(x,l\equ 1;\tilde{t}) 
G_{\tilde\lambda}(l\equ 1,y;t-\tilde{t})}{G_{\tilde\lambda}(x,y;t)}
\Big|_{x=y=0}.
\eeq
We are interested in the behaviour of this expression in the
thermodynamic limit, which is tantamount to letting the cosmological
constant approach its critical value,
$\tilde{\lambda}\to \tilde{\lambda}_c$. In this limit,
\rf{2.11} yields simply a constant,
$\tilde{\Omega}_1^{\rm norm} \buildrel{a\rightarrow
0}\over\longrightarrow 2$. This is a general feature of 
configurations with one or several pinches. For example, generalizing
to geometries with a single pinch of length $l$ gives a coefficient
$2l$ in the large-volume limit. As an example of a more complicated
configuration, the normalized coefficient for
histories with one pinch of length $l_{1}$ and a second one of
length $l_{2}$ becomes in this limit\footnote{Let us take the
opportunity to correct some misprints in equation (29) of \cite{al},
which has been used in deriving 
formula \rf{2.12}. The correct equation reads
$$
G_\l(l_1,l_2;t) = \frac{F^{2t} (1-F^2)^2B_t^{l_1+l_2}}
{l_2 B_t^2 A_t^{l_1+l_2}}
\sum_{k=0}^{\min (l_1,l_2)-1} \frac{(l_1+l_2-k-1)!}{(l_1-k-1)!(l_2-k-1)!k!}
\left(- \frac{A_tC_t}{B_t^2}\right)^k,
$$
where $F,A_t,B_t$ and $C_t$ are defined in \cite{al}.}
\beq{2.12}
\sim u^{l_{1}+l_{2}}:\hspace{0.6cm}
\buildrel{a\rightarrow 0}\over\longrightarrow\,
3\sum_{k=0}^{{\rm min}(l_{1},l_{2})-1}(-1)^{k} 
\frac{(l_{1}+l_{2}-k-1)!}{(l_{1}-k-1)! (l_{2}-k-1)! k!}.
\eeq

By contrast, let us now calculate the first bulk contribution, which
occurs at order $u^{3}$. The contribution to $\tilde{\Omega}_3$
is simply $N$, from counting the number of triangle graphs in the
2d geometry. 
Taking the Laplace transform, we obtain
\beq{2.13}
\sum_{N} \e^{-\tilde{\lambda} N} N \equiv
-\frac{\partial}{\partial \tilde{\lambda}} G(\tilde{\lambda},t)
\equiv \la N \ra\, G(\tilde{\lambda},t). 
\eeq
Evaluating the expectation value of $N$ in the continuum limit,
one finds
\bea
\la N \ra\ &=&- G(\tilde{\lambda},t)^{-1}\, 
\frac{\partial}{\partial \tilde{\lambda}} 
G(\tilde{\lambda},t)\\
\buildrel{a\rightarrow 0}\over\longrightarrow
&&-\frac{4\, (1-\e^{-2 T\sqrt{\Lambda}}-T\sqrt{\Lambda} 
(1-\e^{-2 T\sqrt{\Lambda}}))}{a^{2}\Lambda 
(1-\e^{-2 T\sqrt{\Lambda}})}\,
\buildrel{T\;{\rm large}}\over\longrightarrow\,
\frac{4 T}{a^{2}\sqrt{\Lambda}}.\label{2.14}
\eea
(We are using the notation of \cite{al}, with $T$ and $\Lambda$
the continuum length of the two-geometry in `time'-direction and
the renormalized cosmological constant.)
This diverges exactly the way one would expect from a volume term.
It reiterates the conclusion of \cite{al} that all macroscopic
metric variables
scale canonically in the Lorentzian gravity model. 

We conclude that in the thermodynamic limit, pinching terms will
be suppressed since their number is proportional to $N^{0}$,
whereas the (connected) bulk diagrams behave like $\sim N^{1}$.
For large $N$, the pinch contributions must therefore factorize
according to
\beq{2.15}
(1+\sum_{n\geq 1} \tilde{\Omega}_n (T) u^n )=
(1+N^{0} \sum_{m\geq 1} p_m (T) u^m )\,
(1+N\, \sum_{n\geq 1} \tilde{\omega}_n (T) u^n +O(N^{2})).
\eeq 
Taking the logarithm, we see that the sum
$ (1+ \sum p_m (T) u^m )\sim N^{0}$ 
will only contribute a constant term to the free energy, which
does not affect the universal behaviour of the model.
We will make no attempt to calculate it explicitly. Similar
considerations apply to the high-$T$ expansion of the
magnetic susceptibility in the presence of gravity. The pinch
contributions factorize, and we will only need to compute the 
multiplicity $\tilde{\omega}_{n}^{(2)}$
of bulk polygon graphs with two odd vertices per triangle in
\beq{2.16}
\chi\sim (1+\sum_{n\geq 1}\tilde{\omega}_{n}^{(2)}u^{n}).
\eeq

\begin{figure}
\centerline{\hbox{\psfig{figure=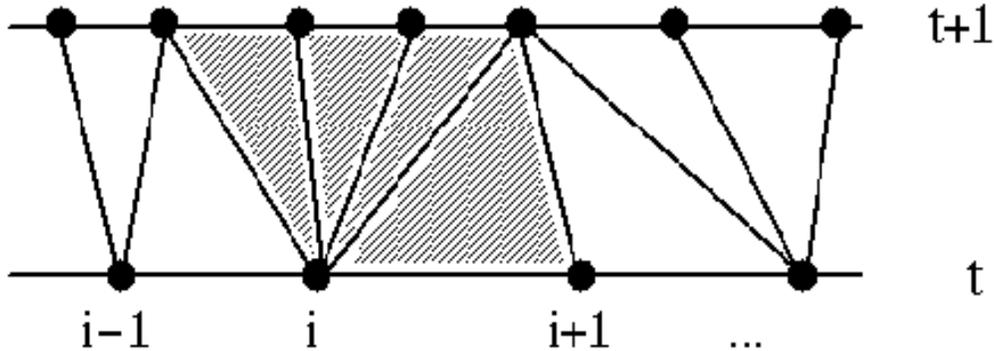,height=6cm}}}
\caption[fig2]{The triangles contributing to the weight at the
vertex $i$.}
\label{fig2}
\end{figure} 

Our next step will be to derive the probability distribution of
the coordination numbers in the Lorentzian gravity model, in the 
thermodynamic limit as
the cosmological constant $\l \to \l_c \equ \ln 2$. 
Recall that when generating an interpolating space-time between
an initial and a final spatial geometry, the geometry of 
each space-time `sandwich' with $\Delta t\equ 1$ is independent of the
previous one in the sense that there are no local constraints
on how the numbers $k_i\geq 1$ of time-like future-pointing links 
can be chosen at each vertex $i$ \cite{al}. Having reached a spatial
slice at time $t$, we can generate the space-time between 
$t$ and $t\pl 1$ proceeding from `left to right'.
To each vertex $i$ at time $t$ we associate $k_i$ time-like
links (ending at vertices of the subsequent spatial slice at $t+1$) 
and the space-like link to the right of the vertex. 
There are therefore exactly $k_i$ triangles associated with the
vertex $i$, contributing with a weight factor $e^{-k\l}$ to the 
action, as illustrated in Fig.\ \ref{fig2}. Since
the assignment of the order $k_i$ of outgoing time-like links to the 
vertex is completely independent of the $k$-assignments of 
other vertices, the probability distribution for $k$ outgoing
future-directed links is given by 
\beq{2.19}
p_{\l}(k) = e^{-k\l} (e^{\l}-1). 
\eeq
Strictly speaking, the argument leading to eq.\ \rf{2.19} is only correct
in the continuum limit in which `extreme pinching' to vanishing 
spatial length $l\equ 0$ does not occur (for off-critical $\l$, 
relation \rf{2.19} must be modified to account for the fact that moves
changing the torus topology are forbidden). Fortunately this is the only 
case we are interested in, and the final probability distribution is
therefore obtained by setting $\e^{-\l}=1/2$ in \rf{2.19}, yielding
\beq{2.20}
p(k)\equiv p_{\l_{c}}(k) = \frac{1}{2^k}.
\eeq
For reasons of symmetry, the distribution of incoming time-like
links at $i$ (originating at the slice at $t\mi 1$) is of course identical.
Given relation \rf{2.20}, we can now compute the 
probability distribution $\tilde{p}(j)$ of
the vertex order, i.e. of the 
{\it total} number $j$ of links meeting at a vertex 
(incoming {\it and} outgoing time-like and space-like links),
\beq{2.21}
\tilde{p}(j)=\frac{j\mi 3}{2^{j-2}},\hspace{1cm} j\geq 4.
\eeq

With the distribution \rf{2.20} in hand, we can now embark on the actual
counting of diagrams contributing to the susceptibility coefficients 
$\tilde{\omega}_{n}^{(2)}$ in \rf{2.16}. 
We will only quote the results up to order $n\equ 5$. Further details
of the counting procedure will appear elsewhere. The average numbers
of diagrams per triangle (i.e. per unit volume) are listed in the
table below.

\begin{center}
\vspace{.5cm}
\begin{tabular}{|r||r|r|r|r|}
\hline
{\rule[-3mm]{0mm}{8mm} $n$} & open & closed & disconnected & total\\ 
\hline\hline  
{\rule[-3mm]{0mm}{8mm} 1} & $\frac{3}{2}$ & 0 & 0 & $\frac{3}{2}$ \\ 
 \hline 
{\rule[-3mm]{0mm}{8mm} 2} & $8 \frac{1}{2}$ & 0 & 0 & $8 \frac{1}{2}$ \\ 
\hline 
{\rule[-3mm]{0mm}{8mm} 3} & $43 \frac{1}{2}$ & 0 & 0 & $43 \frac{1}{2}$ \\ 
 \hline 
{\rule[-3mm]{0mm}{8mm} 4} & $214 \frac{5}{6}$ & 14 & $-17$ & $211 
\frac{5}{6}$\\ \hline 
{\rule[-3mm]{0mm}{8mm} 5} & \hspace{.8cm} $1038 \frac{1}{6}$ & 
\hspace{.8cm} $134 \frac{5}{18}$ & 
$-174 \frac{17}{18}$ & \hspace{.8cm} $997 \frac{1}{2}$\\  \hline
\end{tabular}
\vspace{.5cm}
\end{center}

\noindent
Open graphs are connected graphs without any self-intersections.
Closed graphs are connected graphs which are not open.
The disconnected graphs consist of two or more components and
contribute with a minus sign.

In order to double-check our results at order 4 and 5, where the
counting becomes slightly involved, we have performed a numerical
check on the coefficients $\tilde{\omega}_{n}^{(2)}$ listed above. 
This was done by computer-generating histories of
length $\Delta t\sim 100$, with an initial spatial slice of length
$\Delta l\equ 200$, and counting diagrams of a given type.
The results are given in the table below and in very good agreement with the
exact calculation. They are based on a total of $\sim 3\times 10^{5}$
vertices at order 4 and $\sim 9\times 10^{5}$ vertices at order 5.
We have not listed the counting of disconnected
graphs separately, since it follows closely the counting of 
closed connected graphs.

\begin{center}
\vspace{.5cm}
\begin{tabular}{|r||c|c|}\hline
{\rule[-3mm]{0mm}{8mm} $n$} & open & closed \\ 
\hline\hline  
{\rule[-3mm]{0mm}{8mm} 4} & $214.642 \pm 0.179$ & $13.996 \pm 0.007$\\ \hline 
{\rule[-3mm]{0mm}{8mm} 5} & \hspace{.4cm} $1037.770 \pm 0.751$ 
\hspace{.4cm} & 
\hspace{.4cm} $134.197 \pm 0.098$ \hspace{.5cm}\\  \hline
\end{tabular}
\vspace{.5cm}
\end{center}

\subsection{Evaluation of results}

In order to evaluate the results from the high-$T$ expansion, we
assume a simple behaviour of the susceptibility of the
form 
\beq{4.1}
\chi (u)\sim A\, \Bigl( 1-\frac{u}{u_c}\Bigr)^{-\gamma}+B
\eeq
near the critical point $u_c$, with analytic functions $A$ and $B$.
Using the ratio method
(see, for example, \cite{guttmann}), we have fitted the 
susceptibility coefficients to 
\beq{4.2}
r_{n}=\frac{\tilde{\omega}_n^{(2)}}{\tilde{\omega}_{n-1}^{(2)}}=
\frac{1}{u_c} \Bigl( 1+\frac{\gamma \mi1}{n}\Bigr).
\eeq
Plotting the ratios $r_{n}$ linearly against $1/n$ for
$n\in 1\ldots n_{\rm max}$, we have
extracted the following estimates for the critical point
$u_c$ and the critical susceptibility exponent $\gamma$:
\begin{center}
\vspace{.5cm}
\begin{tabular}{|c||c|c|}\hline
{\rule[-3mm]{0mm}{8mm} $n_{\rm max}$ }  
& critical point & critical 
exponent \\ \hline\hline  
{\rule[-3mm]{0mm}{8mm} 3} & \hspace{.2cm} $u_c=0.2488$ 
\hspace{.2cm} & 
$\gamma =1.820$\\ \hline
{\rule[-3mm]{0mm}{8mm} 4} & $u_c=0.2462$ & 
$\gamma =1.789$\\ \hline 
{\rule[-3mm]{0mm}{8mm} 5} & $u_c=0.2458$ & 
$\gamma =1.783$\\  \hline
\end{tabular}
\vspace{.5cm}
\end{center}

The estimates for the critical exponent should be compared to
the exact values for $\g$ for the Ising model on a fixed, regular
lattice and on dynamically triangulated lattices (Ising spins
coupled to Euclidean quantum gravity), which are $\g^{\rm reg}\equ 1.75$
and $\g^{\rm dt}\equ 2$ respectively.
The data from the high-$T$ expansion clearly favour 
$\g =1.75$ in our model. Indeed, the estimates for $\g$ are remarkably
close to this value, given that we are working only up to order 5
in the expansion parameter $u\equ\tanh \b$. 
The conclusion that the critical exponents of the Ising model coupled
to Lorentzian quantum gravity coincide with those found on regular
lattices is also supported by the 
Monte Carlo simulations we have performed.

\begin{figure}
\centerline{\hbox{\psfig{figure=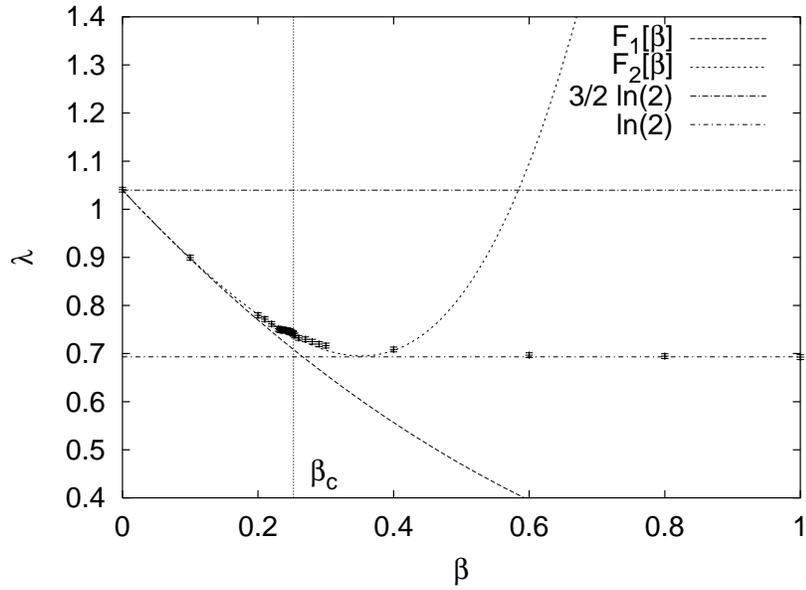,height=8cm,angle=-90}}}
\vspace{36pt}
\centerline{\hbox{\psfig{figure=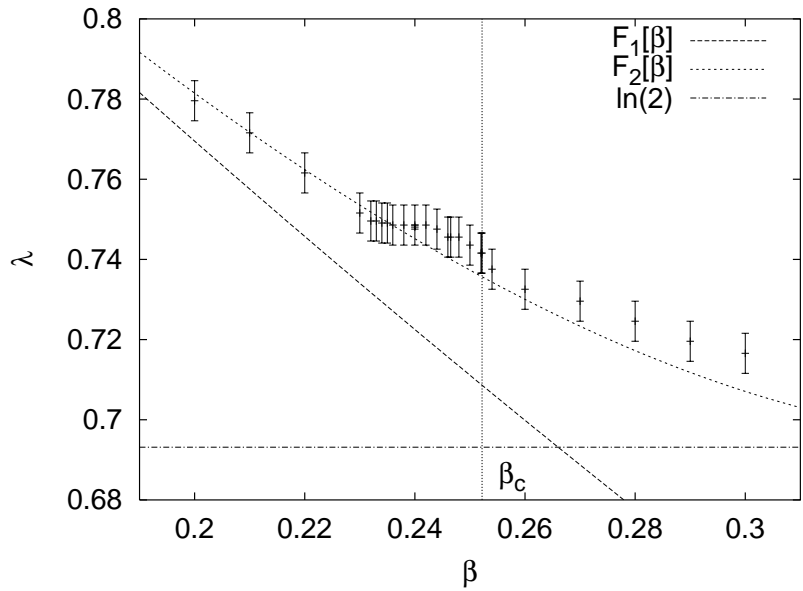,height=8cm,angle=-90}}}
\caption[fig3]{The critical cosmological constant as a function of the
Ising coupling $\beta$, as measured
by Monte Carlo simulations ($t\equ 32$, $N\equ 2048$), 
and compared to the corresponding
high-$T$ expansions $F_1(\b)$ and $F_2(\b)$ at order 0 and order 6.}
\label{fig3}
\end{figure}

However, before turning to a detailed
description of the simulations we would like to illustrate how well 
the high-$T$ expansion works even at this rather low order.
We will compare the $\b$--dependent 
cosmological constant $\l_c(\b)$ defined in eq.\ \rf{add2.6}, which can
be measured directly in the Monte Carlo simulation, with the same
quantity obtained from the high-$T$ expansion. 
Recall that in the thermodynamic limit 
$\l_c(\b)$ is essentially given by the spin free energy, 
eq.\ \rf{add2.5},
which can be computed in the small-$\b$ expansion.
We have determined the density $\tilde{f}(u)$,
defined in eq.\ \rf{add2.8}, by counting closed
polygon graphs in the high-$T$ expansion up to order 6. 
Inserting this into formula \rf{add2.6} leads to
\beq{add2.sc}
\l_c^{\rm high-T}(\b) = \l_c +\oh \ln 2 + \frac{3}{2}\ln \cosh \b + 
u^3 +\frac{5}{3} u^4 +\frac{35}{9} u^5 +\frac{263}{27} u^6.
\eeq
In Fig.\ \ref{fig3} we show the data points
for $\l_c(\b)-3\b/2$ as measured
by the Monte Carlo simulation\footnote{The subtraction of $3\b/2$ 
has been performed to ensure a finite limit as $\b \to \infty$. 
It corresponds to using the action 
$\b \sum_{< ij >} (\sg_i\sg_j-1)$
in eq.\ \rf{2.1}, whose ground state has energy zero rather 
than $-3 \b N/2$.}. Since $\l_c \equ \ln 2$ in pure gravity,
the data should approach $\frac{3}{2}\ln 2$ for
$\b \to 0$ and $\l_c \equ\ln 2$ for $\b \to \infty$, both of which
are well satisfied. 
In order to quantify the effect of the $u$-expansion, 
we have plotted both the zeroth-order expression 
$F_1(\b) =  \l_c +\oh\ln 2+\frac{3}{2}\ln \cosh \b - \frac{3}{2}\b $,  
and the improved sixth-order expression 
$F_2(\b) = \l^{\rm high-T}_c(\b) -\frac{3}{2}\b$. The latter
agrees well with the measured Monte Carlo values right up to the 
neighbourhood of the critical Ising coupling 
$\b_c$. At the critical point $\b_c$ the measured function $\l_c(\b)$ 
exhibits a cusp. This reflects the singular part 
contained in $\l_c(\b)$ which of course cannot be captured by 
simply plotting the analytic function \rf{add2.6}. 
   
\section{The Monte Carlo simulation}\label{simulation}

Monte Carlo simulations have been used successfully in the study
of Euclidean 2d quantum gravity. The formalism known as `dynamical
triangulations' provides a regularization of the functional integral 
well-suited for such simulations, allowing in addition 
for a straightforward matter coupling of Gaussian 
fields as well as of spin degrees of freedom. 
Extensive computer simulations of the combined 
gravity-matter systems have been performed, leading to results 
in perfect agreement with exact results derived from Liouville theory 
and matrix model calculations. 

\begin{figure}
\centerline{\hbox{\psfig{figure=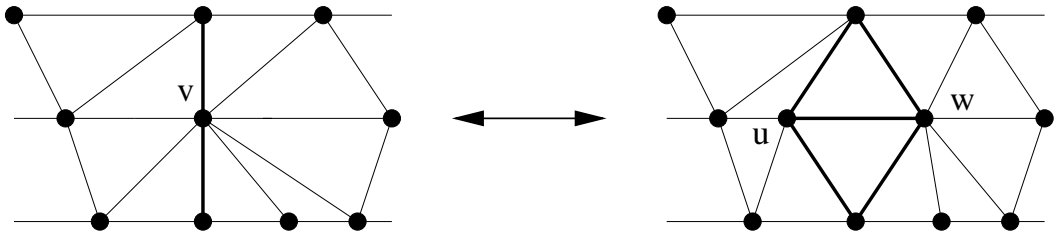,height=3cm,angle=-90}}}
\caption[move]{The move used in the Monte Carlo updating of 
the geometry.}
\label{move}
\end{figure}

The Lorentzian model resembles the dynamically triangulated model 
in that its dynamics is associated with the fluctuating connectivity
of the triangulations contributing to the path integral. 
This allows us to take over many of the techniques from the computer 
simulations of the dynamically triangulated models. 
We must specify the update of both the geometry and the
matter fields, the latter being standard: for a given triangulation we 
update the spin configurations 
by the same spin cluster algorithms used for dynamical triangulations.
This presents no problems since our configurations form a subset 
of the full set of dynamical triangulations used in Euclidean 
quantum gravity (on the torus). During the update of geometry, we want 
to keep the number of time-slices fixed while allowing any space-like 
fluctuations compatible with the model. A local change of geometry or
`move' which is clearly ergodic (i.e.\ can generate any of the 
allowed configurations when applied successively) is shown in 
Fig.\ \ref{move}. 
It consists in deleting the two triangles adjacent to 
a given space-like link (if the resulting configuration is allowed).
Its inverse is a `split' of a given vertex and two neighbouring time-like 
links into two, thereby creating a new space-like link, as well 
as two new triangles. 
This is a special case of the so-called `grand canonical move' sometimes 
used in dynamical triangulation simulations \cite{adfo,jkp,book}, and 
does not preserve the total volume of space-time. 

Detailed balance equations for the move can be derived 
from standard considerations \cite{book}. Let $N_V$ denote 
the number of vertices ($N_V = N/2$, where $N$ is the 
number of triangles), and $v$ a specific vertex.
For pure Lorentzian gravity without matter,
the equation for detailed balance reads
\beq{mc1}
P(N_V) \frac{ P(N_V \!\to\! N_V\!+\!1)}{N_V k_{in} k_{out}}=
P(N_V\!+\!1) P(N_V\!+\!1\! \to \!N_V),
\eeq
where $P(N_V)= \frac{e^{-2\l N_V}}{N_V !}$ is the probability
distribution for labelled triangulations, and $k_{in}$ and $k_{out}$
count the incoming and outgoing time-like links at $v$ (see Fig.\
\ref{move}).  We are still free to choose $P(N_V \!\to\! N_V\!+\!1)$
and $ P(N_V\!+\!1\! \to \!N_V)$ such that condition \rf{mc1} is
satisfied.  Once a transition probability $ P(N_V\!+\!1\! \to \!N_V)$, 
say, has
been chosen, it will be tested during the simulation against the
uniform probability distribution between 0 and 1 as follows. Choose a
random number $r \in ]0,1]$. Then, if the move is {\it allowed} (i.e.\
if the resulting triangulation belongs to the allowed class of
configurations) it is accepted if $P(N_V\!+\!1\! \to \!N_V) > r$.  If
it is not allowed, one proceeds to the next move.

It is straightforward to generalize the updating of geometry to
include Ising spins. The spin Hamiltonian is included in $P(N_V)$,
which now becomes a function of both $N_V$ and the spin
configurations. When inserting a vertex $v$, one has
to specify at the same time a spin associated with $v$. 
The choice of spin up or down
is made with probability 1/2, and the final result tested as in the
case of the pure geometry update.

We have performed the computer simulation for surfaces with 
toroidal topology and for system sizes of $N$=2048, 4050, 8192,
16200 and 32768 triangles, and with a number $t$=32, 45, 64, 90 and
128 of time-slices respectively. Since the moves are not
volume-preserving, fixing the system size to $N$ is implemented
as follows: we allow the volume to fluctuate within a certain, not too
wide range, and collect for every sweep the first
configuration with volume $N$. The volume 
fluctuations are controlled by adding a term
$\delta\lambda\, (\Delta N)^2$ to the action, where $\Delta N$ is the
deviation of the volume from its desired value $N$. This term does not
affect the ensemble of configurations collected, since for all of
them $\Delta N\equ 0$. We find that $\la \Delta N \ra^{-1}\sim
\sqrt{\delta\lambda}$. Finally, one checks that the results obtained do not
depend on the chosen, allowed range of volume fluctuations.  
A sweep is a set of approximately $N_V$ accepted moves.
For each $\beta$-value used in the multi-histogramming analysis
we perform $1.25\times 10^6$ sweeps ($0.75-1.00\times 10^6$ for $N=32768$). 
Measurements are made every $5$ sweeps and errors are computed by 
data binning.

\subsection{Numerical results for the spin system}

The determination of the critical properties of the 
Ising spin system coupled to Lorentzian gravity
proceeds in two steps (see \cite{abt} for a recent, more 
complete discussion in the context of 2d Euclidean quantum gravity). 
We first locate the critical $\b$-value where the system undergoes 
a transition from a magnetized (at large $\b$)
to an unmagnetized phase. Next, we perform simulations
in the neighbourhood of the critical value $\b_c$ and 
use finite-size scaling to determine the critical exponents.
Finite-size scaling is also very useful for determining 
the location of the critical coupling $\b_c$ itself, since
a number of standard observables show a characteristic behaviour for
$\b$ close to $\b_c$.  
The following are some of the observables we have used, together with their
expected finite-size behaviour (see \cite{abt} for a full list): 
 
\beq{chi}
\chi = N (\la m^2\ra -\la |m| \ra^2) \sim N^{\g/\n d_H}~~~~~~~~~~~~
{\rm (susceptibility)}
\eeq
\beq{lnm}
D_{\ln |m|}= N \Big( \la e \ra - \frac{\la e |m|\ra}{\la |m| \ra} \Big)
\sim N^{1/\n d_H}~~~~~~(D_{\ln |m|} \equiv \frac{d \ln |m|}{d \b})
\eeq
\beq{lnm2}
D_{\ln m^2} = N \Big(\la e \ra - \frac{\la e m^2\ra}{\la m^2 \ra} \Big)
\sim N^{1/\n d_H} ~~~~~(D_{\ln m^2} \equiv \frac{d \ln m^2}{d \b}), 
\eeq
where $\g$ and $\n$ are the critical exponents of 
the susceptibility and of the divergent spin-spin correlation length, 
and $d_H$ is the Hausdorff or fractal dimension of space-time. 
For a flat space-time (where of course $d_H\equ d\equ 2$), we have 
$\n d$ = 2, whereas for 
the Ising model coupled to Euclidean quantum gravity $\n d_H$ = 3
(and $d_H \approx 4$). The internal energy density $e$ 
and the magnetization $m$ of the spin system are defined by
\beq{e-m}
e = \frac{-1}{N Z_N(\b)}\frac{d Z_N(\b)}{d\b},~~~~m = 
\frac{1}{N Z_N(\b,H)}\frac{d Z_N(\b,H)}{dH}\; \Big|_{H=0}.
\eeq

In order to find the critical point $\b_c$, we can use the fact that
the pseudo-critical coupling $\b_c(N)$ at volume $N$ is expected to
behave like 
\beq{200}
\b_c(N) \sim \b_c + \frac{c}{N^{1/\n d_H}}
\eeq
close to $\b_c\equ\b_c(N\equ\infty)$, with $c$ a constant.
The observables \rf{chi}-\rf{lnm2} all have well-defined peaks
which we used for a precise location of $\b_c(N)$,
with the help of multi-histogram techniques. Eq.\ \rf{200}
can now be used to extract $\b_c$ and $1/\n d_H$. 
However, it is advantageous to determine first 
$1/\n d_H$ from the peak values of \rf{lnm} and \rf{lnm2}, 
and then substitute this value into \rf{200}, thus 
reducing the number of free parameters in the fit.
Afterwards, one can check that consistent values 
for $1/\n d_H$ are obtained from the observables 
\rf{lnm} and \rf{lnm2} at $\b_c$, using multi-histogramming. 
In our simulations, $1/\n d_H$ extracted from the 
peaks was so close to the Onsager value 1/2 
that we did not hesitate to use $1/\n d_H\equ 1/2$ in \rf{200}. Leaving
it as a free parameter one obtains consistent results, but the error
in $\beta_c$ becomes larger.
Finally, we have determined $\g/\n d_H$ from \rf{chi}, both 
from the peak values and at $\b_c$  
(and again using multi-histogramming). 

In the table below we have listed the $\b_c$-values extracted
from measuring $D_{\ln |m|}$ and $D_{\ln m^2}$, and assuming 
that $1/\n d_H$ = 1/2. For comparison with the high-$T$ results,
we also give the corresponding critical values for the expansion
parameter $u$. Similar
results are obtained from the rest of the observables we have measured.

\vspace{12pt}
\begin{center} 
\begin{tabular}{|c||c|c|}\hline
{\rule[-3mm]{0mm}{8mm} Observable} & $\b_c$, using 
$\frac{1}{\n d_H} = \frac{1}{2}$ &
$u_c= \tanh \b_c$ \\ \hline\hline
{\rule[-3mm]{0mm}{8mm}$D_{\ln |m|}$} & 0.2522 (2)& 0.2470 (2) \\ \hline
{\rule[-3mm]{0mm}{8mm}$D_{\ln m^2}$} & 0.2520 (1)& 0.2468 (2)\\ \hline
\end{tabular}
\end{center}
\vspace{12pt}

\noindent
Column 1 of the following table contains the values of $1/\n d_H$ 
extracted from the peak position for all three observables \rf{chi}--\rf{lnm2} 
by using relation \rf{200} (with free parameters $\beta_c$, $1/\n d_H$ and 
$c$). Column 2 and 3 give $1/\n d_H$ extracted directly from 
\rf{lnm} and \rf{lnm2} by using the peak values of the observables
and their values at $\b_c$.

\vspace{12pt}
\begin{center}
\begin{tabular}{|c||c|c|c|}\hline
{\rule[-3mm]{0mm}{8mm}Observable}  
& $\frac{1}{\n d_H}$ from peak pos.& 
\hspace{.3cm}$\frac{1}{\n d_H}$ at peak \hspace{.3cm} & 
\hspace{.3cm} $\frac{1}{\n d_H}$ at $\b_c$ \hspace{.3cm} 
\\ \hline\hline
{\rule[-3mm]{0mm}{8mm} $\chi$} & 0.52 (2) & ---& ---\\ \hline
{\rule[-3mm]{0mm}{8mm} $D_{\ln |m|}$} & 
0.47 (2)& 0.531 (2) & 0.521 (3)  \\ \hline
{\rule[-3mm]{0mm}{8mm} $D_{\ln m^2}$} 
& 0.53 (1) & 0.531 (1) & 0.520 (3) \\ \hline
\end{tabular}
\end{center}

\vspace{12pt}
\noindent
Lastly, we give the value of $\g/\n d_H$ extracted from the
susceptibility \rf{chi},

\vspace{12pt}
\begin{center}
\begin{tabular}{|c||c|c|}\hline
{\rule[-3mm]{0mm}{8mm} Observable} & value at peak& value at $\b_c$ 
\\ \hline\hline
{\rule[-3mm]{0mm}{8mm}$\chi$} & 0.883 (1) & 0.899 (2)\\ \hline
\end{tabular}
\end{center}

\vspace{12pt}
Comparing the estimates for $\b_c$ from the high-$T$ expansion and
the Monte Carlo simulation, one finds good agreement. 
The results of the simulation strongly indicate
that the critical exponents are given by the Onsager values
$\n d_H \equ 1/2$ and $\g\equ 1.75$. Again, this corroborates 
the conclusion already reached by means of the high-$T$ expansion.
Further evidence that the system belongs to the Onsager and not
the Euclidean gravity universality class comes from measuring the
magnetization exponent $\b_m/\n d_H$ and the specific heat exponent
$\alpha/\n d_H$. 
Their Onsager values are $1/16$ and $0$, whereas in Euclidean gravity 
they are $1/6$ and $-1/3$. 
In our model, the magnetization exponent determined from $\la |m|
\ra_{\b=\b_c}\sim N^{-\b_m/\n d_H}$ was found to be 
$\b_m/\n d_H = 0.070(1)$, favouring the Onsager value 0.0625
over the Euclidean gravity value $0.166\bar{6}$. 
The specific heat exponent was obtained from the finite-size scaling
of the values of the specific heat peaks $C_V\sim N^{\alpha/\n
d_H}$. A power fit yields $\alpha/\n d_H=0.0861(7)$ at $\chi^2/{\rm
dof}=11.6$ whereas a logarithmic fit gives $\chi^2/{\rm dof}=1.57$, 
supporting the conjecture that $\alpha\equ 0$.

We do not have independent measurements of the critical parameters 
$\n$ and $d_H$ from the spin sector alone, but we will determine the 
Hausdorff dimension $d_H$ in the next subsection from an
analysis of the geometry of Lorentzian quantum gravity coupled to
Ising spins.

\subsection{Numerical results for the geometry}

As is well-known from both analytical studies \cite{analytic,ajw,aaa}
and numerous Monte Carlo simulations (\cite{numerical,ajw} and
references in \cite{book}), finite-size scaling is a powerful tool for
determining the fractal space-time structure of
two-dimensional Euclidean quantum gravity. 
The same technique can be used to investigate the geometric
properties of two-dimensional Lorentzian quantum gravity.

In a given triangulation, we define the distance between two vertices 
as the minimal length of a connected path of links between them.
In 2d Euclidean quantum gravity this notion of distance 
becomes proportional to the true geodesic
distance between the vertices in the infinite-volume limit.  
We will assume this is also true for the present model. 
All diffeomorphism-invariant correlation
functions of matter fields must be functions of this geodesic
distance. Both the geodesic distance and the fractal dimension 
appear in the expression for the volume 
\beq{300}
N(r) \sim r^{d_H} ~~~{\rm for}~~~ r \ll N^{1/d_H},
\eeq
which denotes the number of vertices (or triangles)
inside a ball (a disc in dimension 2) of link-radius $r$. 
If $n_v(r)$ denotes 
the number of vertices at distance $r$ from a fixed vertex $v$, 
relation \rf{300} implies that
\beq{301}
n_v(r) \sim r^{d_H-1} ~~~{\rm for}~~~ r \ll N^{1/d_H} .
\eeq
Finite-size scaling for an observable $A$ then leads to a scaling
of the correlation function integrated over all points at 
distance $r$ from a vertex according to 
\beq{302}
\la A(r) A(0)\ra_N \sim N^{1-1/d_H-\Del_A} F_A(x),~~~~x= r/N^{1/d_H}.
\eeq
The factor $N^{1-1/d_H}$ comes from the integration over points
at distance $r$ from vertex $v$, using \rf{301}, while $\Del_A$ is the genuine
dynamical exponent of the correlator.

\begin{figure}
\centerline{\hbox{\psfig{figure=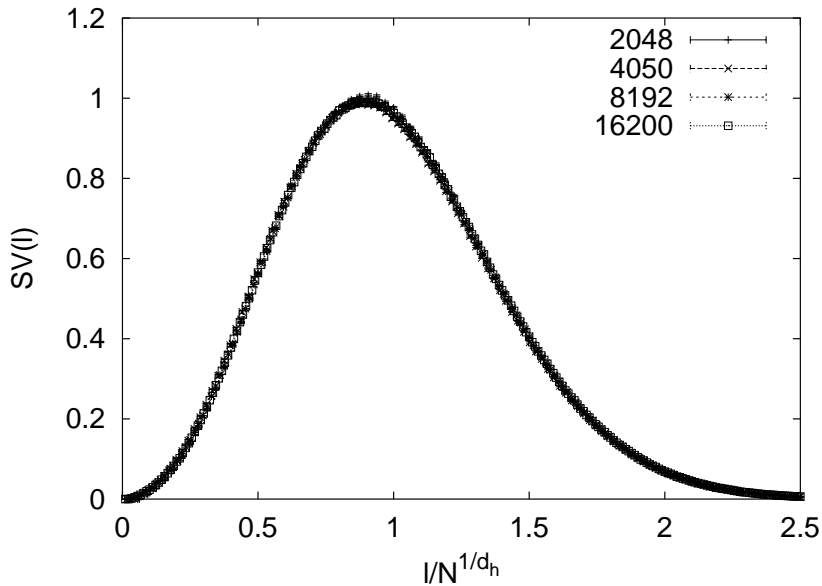,height=8cm,angle=-90}}}
\caption[figSV]{The distribution of spatial volumes $SV(l)$ at 
$\beta\equ\beta_c$, rescaled according to eq. \protect\rf{304}. 
We have set $d_H\equ 2.0$.}
\label{figSV}
\end{figure}

By measuring correlation functions for various volumes $N$, one can
determine $d_H$ and the critical exponents. We will concentrate
here on the Hausdorff dimension $d_H$. 
One first rescales the height of the measured distributions
$\la A(r) A(0)\ra_N$ to a common value. However,
the distributions measured for different $N$ will
still have different width as functions of $r$. By appropriately
rescaling $r$, they can then be made to overlap in a 
single, universal function $F_A(x)$. 
From a technical point of view it is important to work with 
the shifted variable
\beq{303}
x= \frac{r+a_A}{N^{1/d_H}},
\eeq
where the shift $a_A$ may depend on the observable $A$. Using
eq.\ \rf{303} takes into
account in an efficient way the major part of the short-distance
lattice artifacts, as has been discussed carefully in \cite{abj,ajw,aaa}. 
Applying standard procedures from Euclidean 2d quantum gravity then
leads to the results summarized in Table \ref{haus}. 
The observables appearing in Table \ref{haus} are: 
(i) the number $n_v(r)$
of vertices at a given (link-)distance $r$ from a fixed vertex $v$,
which may be
viewed as the correlation function of the unit operator in quantum
gravity \cite{analytic}; 
(ii) the number $s_{up}(r)$ of spins at distance $r$ from a vertex $v$
which are aligned with the spin at $v$; 
(iii) the number $s_{down}(r)$ of spins
with orientation opposite to the spin at $v$; (iv) the
spin-spin correlation function $s(r)$ between vertices 
separated by a geodesic distance $r$;
(v) the function $S(r)$, obtained by 
integrating $s(r)$ over all vertices at distance $r$ from a vertex $v$;
(vi) the distribution $SV(l)$ of spatial volumes,
with $l$ denoting the length of a given time-slice. 
For the shift $a_A$, we obtained the estimate 
$-4<a_A<-1$. Unfortunately our statistics 
was not good enough to determine it with more accuracy. 
However, the fact that it is non-vanishing justifies its introduction
in the first place. After a
suitable normalization, we expect the volume distribution to behave like
\beq{304}
SV(l) \sim f(l/N^{1/d_H}).
\eeq
Fig.~\ref{figSV} demonstrates clearly that for the Ising model at 
$\beta\equ \beta_c$, $SV(l)$ scales as anticipated when we set $d_H\equ 2.0$.
Scaling the Ising distributions at $\beta\equ\beta_c(N)$, 
the pseudo-critical point of the magnetic
susceptibility, or considering pure gravity leads to similar results. 

\begin{table}[ht]
\begin{center}
\begin{tabular}{|c||c|c|c|}
\hline
\multicolumn{1}{|c|}{}&
\multicolumn{3}{|c|}{\rule[-3mm]{0mm}{8mm} {$d_H$} }\\ 
\multicolumn{1}{|c|}{\rule[-3mm]{0mm}{8mm} {Observable} } &
\multicolumn{1}{c}{$c=\frac{1}{2}$, $\b=\b_c$}&
\multicolumn{1}{c}{$c=\frac{1}{2}$, $\b=\b_c(N)$}&
\multicolumn{1}{c|}{\hspace{.2cm} $c=0$\hspace{.2cm}}\\
\hline\hline
{\rule[-3mm]{0mm}{8mm} $n_v$ }     & 2.00(5) & 2.00(5) & 2.03(4) \\ 
{\rule[-3mm]{0mm}{8mm} $s_{up}$ }  & 1.92(5) & 2.08(2) & ---\\ 
{\rule[-3mm]{0mm}{8mm} $s_{down}$ } & 2.20(3) & 2.07(2) & ---\\ 
{\rule[-3mm]{0mm}{8mm} $S$  }      & 2.10(3) & 2.20(5) & ---\\ 
{\rule[-3mm]{0mm}{8mm} $s$  }      & 2.05(7) & 2.05(5) & ---\\ 
{\rule[-3mm]{0mm}{8mm} $SV$ }      & 2.00(4) & 2.00(3) & 2.00(3)\\
\hline
\end{tabular}
\end{center}
\caption[haus]{The Hausdorff dimension $d_H$, obtained from
the Ising model scaling at $\b_c$ (column 1), at $\b_c(N)$
(column 2) and from pure gravity (column 3).}
\label{haus}
\end{table}

We conclude from Table \ref{haus} that the Hausdorff dimension of
2d Lorentzian quantum gravity is close to the flat-space value
$d_{H}\equ 2$. This is clearly different from the Euclidean situation
which is characterized by $d_{H}\equ 4$ for pure gravity and
$d_H \geq 4$ in the presence of Ising spins. 
The results for the Lorentzian gravity-matter system are particularly
convincing for the purely geometric observables $n_v(r)$ and $SV(l)$,
which basically coincide with the corresponding measurements obtained
in Lorentzian pure gravity.

\section{Conclusions}

We have presented compelling evidence, coming from a high-$T$
expansion as well as Monte Carlo simulations, that the critical
exponents of the Ising model coupled to Lorentzian gravity are
identical to the exponents in flat space. This is in contrast with the
situation in Euclidean gravity (i.e.\ Liouville gravity), where the
exponents change\footnote{The exponents of the Ising model coupled to
2d Euclidean quantum gravity are equal to those of the 3d spherical
model. It is not understood whether this is a coincidence. More
generally, the exponents of the $(m,m+1)$ minimal conformal model
coupled to 2d Euclidean quantum gravity agree with the critical
exponents of the spherical model in $2m/(m+1)$ dimensions.}.
Similarly, the fractal dimension of space-time is unchanged in the
Lorentzian model after coupling it to a conformal field theory (the
Ising model at the critical point). In Euclidean gravity the fractal
properties of space-time are in general a function of the central
change of the conformal field theory. From the evidence collected so
far, we conclude that matter and geometry couple weakly in Lorentzian
gravity and strongly in Euclidean gravity.

For the case of the Ising model, this difference can be explained in
more detail in geometric terms. As mentioned earlier, it has been
shown in \cite{al} that the difference between Euclidean and
Lorentzian gravity is related to the presence or absence of baby
universes.  On the other hand, it is by now well understood that baby
universes are at the source of the strong coupling between spins and
geometry. This can happen because the spin configuration of a baby
universe can be flipped relative to that of the `parent' universe at
almost no cost in energy since the `baby' and the `parent' are
connected only by a thin tube.  The geometry of two-dimensional
Euclidean quantum gravity is very fractal, with many `pinches' at all
scales, leading to typical spin configurations that look very
different from those on flat space-time.  Moreover, the presence of
Ising spins on the surface effectively enhances the fractal baby
universe structure since it is exactly the lowest energy spin
configurations (apart from the ground state) that involve baby
universes.  The interaction becomes so strong that it tears the
surface apart when we couple more than two Ising spins to the
two-dimensional geometry. This is the origin of the famous $c\equ 1$
barrier of two-dimensional Euclidean quantum gravity.

Once the creation of baby universes is disallowed, as in the case of
the Lorentzian model, the coupling between matter and geometry becomes
weak, and the matter theory has the same critical exponents as in flat
space-time.  This happens although the typical space-time geometry is
by no means flat, a fact we have already emphasized in the
introduction, and which is illustrated by Fig.\ \ref{fluctuate}.  On
the contrary, our model allows for maximal fluctuations of the spatial
volume which can jump from (essentially) zero to infinity in a single
time step.  However, as we have been able to demonstrate, such violent
fluctuations of the two-geometry are still not sufficient to induce a
change in the critical exponents of the Ising model.  From the above
arguments it seems likely that Lorentzian gravity can avoid the $c\equ
1$ barrier. This question is presently under investigation.

\subsection*{Acknowledgements}
We would like to thank C.\ Kristjansen for comments on a preliminary
version of this article.

\end{document}